%% file: 0_main.tex
\documentclass[acmtog]{acmart}
\acmSubmissionID{1654}

\usepackage{booktabs} 

\usepackage[ruled]{algorithm2e} 

\SetAlFnt{\small}
\SetAlCapFnt{\small}
\SetAlCapNameFnt{\small}
\SetAlCapHSkip{0pt}
\usepackage{cleveref}
\usepackage{xspace}
\usepackage{wrapfig}
\usepackage{soul}

\newcommand{\perspectiveM}{\ensuremath{\mathbf{P}}\xspace}
\newcommand{\modelview}{\ensuremath{\mathbf{M}}\xspace}
\newcommand{\humanPerspD}{\ensuremath{\mathbf{D}}\xspace}
\newcommand{\p}{\ensuremath{{\mathbf{p}}}\xspace}
\newcommand{\proj}{\ensuremath{\textrm{proj}}\xspace}

\copyrightyear{2025}
\acmYear{2025}
\setcopyright{rightsretained}
\acmConference[SA Conference Papers '25]{SIGGRAPH Asia 2025 Conference Papers}{December 15--18, 2025}{Hong Kong, Hong Kong}
\acmBooktitle{SIGGRAPH Asia 2025 Conference Papers (SA Conference Papers '25), December 15--18, 2025, Hong Kong, Hong Kong}
\acmDOI{10.1145/3757377.3763902}
\acmISBN{979-8-4007-2137-3/2025/12}

\begin{document}

\title{Capturing Non-Linear Human Perspective in Line Drawings}


\author{Jinfan Yang}
\affiliation{%
	\institution{University of British Columbia}
	\country{Canada}
}
\email{yangjf@cs.ubc.ca}
\author{Leo Foord-Kelcey}
\affiliation{%
	\institution{University of British Columbia}
	\country{Canada}
}
\email{leofk@cs.ubc.ca}
\author{Suzuran Takikawa}
\affiliation{%
	\institution{University of British Columbia}
	\country{Canada}
}
\email{stakikaw@cs.ubc.ca}
\author{Nicholas Vining}
\affiliation{%
	\institution{NVIDIA}
	\country{Canada}
}
\affiliation{%
	\institution{University of British Columbia}
	\country{Canada}
}
\email{nvining@cs.ubc.ca}
\author{Niloy Mitra}
\affiliation{%
	\institution{University College London}
	\country{United Kingdom}
}
\affiliation{%
	\institution{Adobe Research}
	\country{United Kingdom}
}
\email{niloym@gmail.com}
\author{Alla Sheffer}
\affiliation{%
	\institution{University of British Columbia}
	\country{Canada}
}
\email{sheffa@cs.ubc.ca}

\input{01_abstract.tex}

%
%

%
%

\newenvironment{parWithWrapFigure} %
{\begingroup
\setlength{\columnsep}{1em}%
\setlength{\intextsep}{0em}%
\setlength{\arraycolsep}{0pt}} %
{
\endgroup
}


\begin{teaserfigure}
 \includegraphics[width=\linewidth]{figures_comp/teaser_v9.pdf}
  \caption{
  Human sketches~(a, black) use a perspective projection that deviates from the analytical perspective model. We visualize this {\em perspective deviation} by overlaying the sketch over contours projected using best approximating analytic perspective (a, {\color{orange}orange}); sketch and contours shown separately as insets. We learn a model of this human deviation, producing projected contours (b-d, black) that have similar relation to their analytical counterparts. Our deviation model generalizes to other views (c) and shapes (d), maintaining similar relationship between the two sets of contours.}
  \label{fig:teaser}
\end{teaserfigure}

\newcommand{\old}[1]{}
\newcommand{\new}[1]{#1}

\maketitle

\input{1_intro}

\input{2_related}

\input{3_overview}

\input{4_algorithm}

\input{6_results}

\input{7_conclusion}
\input{acks}

\bibliographystyle{ACM-Reference-Format}
\bibliography{ref}

\newpage
\input{figures}

\end{document}

%% file: 01_abstract.tex
\begin{abstract}
Artist-drawn sketches only loosely conform to analytical models of perspective projection; the deviation of human-drawn perspective from analytical perspective models is persistent and well documented, but has yet to be algorithmically replicated. 
We encode this deviation between human and analytic perspectives as a continuous function in 3D space and develop a method to learn it.  
We seek deviation functions that (i)~mimic artist deviation on our training data;  (ii)~generalize to other shapes; (iii) are consistent across different views of the same shape; and (iv)~produce outputs that appear human-drawn. The natural data for learning this deviation is pairs of artist sketches of 3D shapes and best-matching analytical camera views of the same shapes. 
However, a core challenge in learning perspective deviation is the heterogeneity of human drawing choices, combined with relative data paucity (the datasets we rely on have only a few dozen training pairs). 
We sidestep this challenge by learning perspective deviation from an individual pair of an artist sketch of a 3D shape and the contours of the same shape rendered from a best-matching analytical camera view.
We first match contours of the depicted shape to artist strokes, then learn a spatially continuous local perspective {\em deviation} function that modifies the camera perspective projecting the contours to their corresponding strokes. This function retains key geometric properties that artists strive to preserve when depicting 3D content, thus satisfying (i) and (iv) above. We generalize our method to alternative shapes and views (ii,iii) via a self-augmentation approach that algorithmically generates training data for nearby views, and enforces spatial smoothness and consistency across all views.  
 We compare our results to potential alternatives, demonstrating the superiority of the proposed approach.
Code and models will be released upon acceptance. 

\end{abstract}

%% file: 1_intro.tex
\section{Introduction}
\label{sec:intro}

Line drawings, or sketches, are a simple and powerful medium for conveying shapes between humans \cite{eissen08,eissen2011sketching} and have long been hailed as a potential interface for bidirectional human-computer communication \cite{Sutherland}.  Unfortunately, computer generated sketches lack the communication power of human ones, and computers are not yet able to fully parse human sketches~\cite{BessmeltsevLiuSGP2024}. 

When sketching a 3D object (e.g.~\Cref{fig:teaser}a), artists make three key choices: (i)~which  3D curves to draw;  (ii)~what 3D-to-2D projections to employ to project the 3D curves on a 2D medium; and (\old{ii}\new{iii})~how to stylize the 2D projections of the chosen curves.
While the first and last items have been extensively researched (\Cref{sec:related}), the \old{third}\new{second} one~\cite{Hertzmann2024} is less studied and rarely modeled. Computer graphics applications typically use an analytical model of projection (e.g. single vanishing-point perspective or orthographic projection); in contrast, 
artist projection approximates but does \textit{not} match any such analytical model \cite{Pepperell2014,Hertzmann2024}. 
Researchers~\cite{Pepperell2014,Gombrich1951-GOMTSO-3,Singh:2002:FP,Hertzmann2024} speculate that the deviation between artist-employed and analytical perspective models is due to a combination of artists using deliberate distortions as a key mechanism to emphasize essential features and reduce cognitive load, thereby enabling more effective communication; and inherent human imprecision. We refer to this discrepancy between analytical and artist-drawn perspectives as \textit{human perspective deviation}.  

Although this human perspective deviation has been consistently observed, we are aware of only one prior effort to model it. Specifically, Xiao et al.~\shortcite{xiao2022differsketching} note the discrepancy between artist-sketched and analytically projected shape contours, and use a 2D multi-layer perceptron~(MLP) to learn to deform the analytical contours towards their sketched counterparts. However, this approach collectively models all three above-mentioned drawing choices and does not offer a factorized treatment.  Even when applied to contours in their training corpus, their model, trained on professional artist line-drawings,  produces unnatural looking deviations containing high-frequency noise, atypical of artist choices (see \Cref{sec:results}). 

We develop a deviation model that is designed to reflect artist perspective choices.  Since there is no unified perceptual model of artist perspective deviation upon which we can draw,  we propose a learning-based approach and use pairs of artist sketches, dominated by contour strokes, and renders of analytically projected contours of the artist-depicted shapes aligned to match the artist's chosen view as training data  (\Cref{fig:teaser}a).  Given this data, we seek deviation models that (i)~mimic artist deviation on our training data;  (ii)~generalize to other shapes; (iii) are consistent across different views of the same shape; and (iv)~produce outputs that appear human-drawn. 

A core challenge in learning perspective deviation that reflects artist choices is the heterogeneity of human drawing choices, combined with relative data paucity (the datasets we rely on have less than a couple of hundreds of training pairs). Rather than attempting to learn a single unified probabilistic model that can generate outputs that fit individual styles on demand, we learn models that reflect the perspective in individual sketches. While our model can be extended to learning from multiple sketches at once (\Cref{sec:results}). such ``averaged'' models tend to be less expressive. 

We align the renderer camera parameters to best match the artist sketches and algorithmically match the rendered projected contours to artist strokes. We then use these correspondences to learn a {\em perspective deviation} function that maps contours to their matching strokes. Our experiments (\Cref{sec:results}) suggest that generalization across views and shapes requires learning a 3D, rather than a 2D, perspective deviation function.  We therefore model perspective deviation using a 3D spatially varying multiplicative matrix that adjusts the analytical projection matrix; specifically, we use an MLP to define, for every point in 3D space, an associated deviation matrix. To generalize the output perspective deviation across multiple camera views, we use a self-augmentation process where we first learn artistic deviation from the sketch contour pairs above; then, we learn a deviation function across both the original input and synthetic training examples generated from the original deviation.

We train our method on 169 sketch/shape pairs sourced largely from \cite{OpenSketch19} and \cite{xiao2022differsketching}, and show the results of applying these models to different shapes and views throughout the paper and the supplementary material. Our outputs retain the perspective of the input sketches when applied to their corresponding training shapes, both from original and novel camera views (e.g. \Cref{fig:teaser}bc), and translate across shapes (e.g. \Cref{fig:teaser}d). We validate our results both quantitatively and via perceptual studies, and demonstrate their superiority compared to existing and potential alternatives (\Cref{sec:results}).

In summary: our main contribution is to learn human perspective deviation models that capture the characteristics of individual artist choices and that generalize across views and shapes. In the process, we contribute to the understanding of human employed perspective in line drawings, and identify the relevant factors in modeling human perspective deviation.
Beyond addressing the technical challenge of modeling perspective deviation for individual artists and inputs, our approach advances the understanding of how computational models can replicate human perception.

%% file: 2_related.tex
\section{Related Work}
\label{sec:related}

\paragraph{Sketching with Perspective}
Analytical, linear, perspective projection is ubiquitously used for precise depiction of 3D content from a given viewpoint in manually drafted technical drawings and computer generated renders.  
Artists are often encouraged \cite{eissen08} to aim for analytic perspective and historically have attempted to accurately reproduce it; for example, it is speculated that the Dutch masters used {\em camera obscura} to capture this perspective \cite{steadman2002vermeer}.
However, various user studies have demonstrated that artists almost never use precise linear perspective for sketching~\cite{Koenderink2016,Pepperell2014}. Some deviations arise due to faulty estimation~\cite{warwick22265,kubovy1986psychology} while others are the result of artists intentionally using varying (local) perspective~\cite{Hertzmann2024,Pepperell2014,Singh:2002:FP,singh_perspective:09,patrick:perspective:05}. 
Research on human perception of 2D depiction of 3D objects strongly suggests that humans make systematic errors when estimating foreshortened shapes and dimensions even for simple tasks \cite{nicholls1995cube,reith1995projective,taylor1997judgments,Koenderink:91}.
While studies suggest that using scaffolds for guidance~\cite{singh_perspective:09,hennessey:sketch:17} improves the alignment of artist and analytic perspectives, artists often forego scaffolds when sketching free-hand.

\paragraph{Non-Photorealistic Rendering (NPR)}
Numerous NPR methods explore the use of  line drawings for effectively conveying shape, and investigate \textit{which} surface curves or contours to draw, e.g.~\cite{Cole:2008:WDP} and \textit{how} to stylize their 2D projections, e.g. \cite{wang2024instantstyle,hertz2023StyleAligned}.
DeCarlo et al.~\shortcite{DeCarlo:2003:SCF} generate collections of curves that emphasize object features;  recent variants (e.g.,  H\"ahnlein et al.~\shortcite{CAD2Sketch10.1145/3550454.3555488}, Liao et al.~\shortcite{freehand_mechParts_24}) convert CAD sequences to concept sketches, blending geometric precision with stylistic abstractions to emulate human sketches. All above methods explicitly or implicitly rely on traditional analytical perspective. 

Advances in machine learning have opened new possibilities for synthesizing line drawings from 3D shapes, 
including neural style transfer \cite{StyleTransfer7780634}, image-to-image translation \cite{pix2pix2017}, or using CLIP scoring to distill shape abstractions~\cite{vinker2022clipasso}. Liu et al.~\shortcite{Liu_2020_CVPR}  propose a neural framework for generating contour lines directly from 3D models, showcasing the ability of neural networks to learn artistic cues. 
Chen et al.~\shortcite{chan2022drawings} propose neural variants for synthesizing line drawings that simultaneously capture geometric accuracy and semantic meaning. 
These methods demonstrate the potential of machine learning to mimic artistic styles, but are mainly trained on synthetic renderings of 3D models using analytical projections, and learn styles rather than human perspective.
These methods do not address the perspective distortion humans naturally introduce, as particularly evident when comparing NPR outputs with human-drawn sketches
(see supplementary).

Closest to our work, DifferSketching~\cite{xiao2022differsketching} use a data driven approach to learn a 2D difference function between projected contours and artist sketches (assumed to be drawn in the same view) from a large collection of paired contours and sketches. As our experiments show, despite relying on professional sketches, their method introduces notable high-frequency distortion when applied even to their training inputs. Training the method on a subset of the dataset, e.g., drawings of the same shape, or single accurate drawings increases rather than decreases this distortion. We compare theirs with our approach on their dataset in \Cref{sec:results}.

\paragraph{Sketch-Based Modeling}
Sketch-based modeling systems focus on creating 3D models from 2D sketches; see \cite{OLSEN200985,BessmeltsevLiuSGP2024,Choi_2024} for comprehensive surveys. Many such systems ignore the problem of perspective entirely, and use 2D contour curves drawn in the image plane as input \cite{FiberMesh:2007, zhang2022creatureshop, MonsterMash:2020, Li:2018:SketchCNN}; they create 3D geometry by inflating these contours and incorporate depth either by explicit annotation or relying on stroke draw order. These methods implicitly assume orthographic perspective. Other methods rely on sketched input from multiple views, where artists sketch strokes from different viewpoints onto existing 3D geometry (e.g., \cite{Teddy10.1145/311535.311602,kara2007sketch,de2015secondskin}). Several methods require users to manually specify analytic perspective ``scaffolds'' to regularize perspective \cite{schmidt2009scaffold}, or use strokes to define transient surfaces to recover 3D curves \cite{bae2008ilovesketch}.

Works addressing 3D reconstruction from single sketches observe that user inputs have inexact perspective, but seek to correct or sidestep this inexactness by  detecting and enforcing different regularization cues \cite{crossShade:12,true2Form:2014}, construction lines \cite{Lift3DSA20},  or local symmetries \cite{hahnlein2022symmetry}.
Recent developments have shifted towards data-driven approaches
by leveraging 3D datasets, synthetically rendered with a pinhole camera model with either non-photorealistic rendering or manual contour tracing, to create training and test data \cite{liu2023zero1to3, li2022free2cad, LiuSketchDream}. When applied to human sketches, they frequently produce unexpected or inconsistent outputs, highlighting the need for frameworks that explicitly incorporate human perceptual biases.

%% file: 3_overview.tex
\section{Method}
\label{sec:alg}

\subsection{Overview}
\label{sec:overview}

\paragraph{Modelling Human Perspective.}
As discussed in \Cref{sec:related} while artistic perspective typically deviates from analytic one, this deviation is relatively subtle, and changes gradually across drawings, with parts of the content drawn larger or smaller relative to their analytical projection. 
To capture these properties, we model artists' perspective as a 3D deviation from a standard pinhole camera projection that smoothly varies across 3D space (we treat orthographic projection as a special case of perspective with the camera placed at infinity).

\begin{figure*}[t!]
    \centering
    \includegraphics[width=\linewidth]{figures_comp/overview_v10.pdf}
    \caption{Algorithm overview: Given an input sketch (black) and corresponding analytically projected 3D shape contours ({\color{orange}orange}) in a matching view (a), we match the  contours to artist strokes (b). We then use a two-step process to learn the deviation between the contour and sketch projections (c-e): we learn an initial deviation function $\humanPerspD(\p)$ that balances satisfying the computed matches against adherence to core deviation properties and apply the learned deviation to the input contours (c); we augment our learning data with synthetic sketch/contour pairs and re-learn a deviation that best fits the augmented training set (d,e). The contours projected using our deviation align with the artist's strokes (f). Our deviation consistently generalizes to other views (g).}
    \label{fig:overview}
    \vspace{-.15in}
\end{figure*}

\paragraph*{Setup}
In computer graphics, projection is handled analytically through the camera projection matrix \perspectiveM and the modelview matrix \modelview. For simplicity we refer to the product of these matrices $\mathbf{C} =\perspectiveM \modelview$ as the camera projection matrix.  
When applying a perspective projection, any point $\p \in \mathbb{R}^3$ on a 3D shape is mapped, working in homogeneous coordinates, to $\p \rightarrow \mathbf{C} [\p;1]$.  
Then, $\mathbf{C} [\p;1]$ is converted to 2D points in image space by performing a perspective divide (see \cite{foley1996computer}). 

We model human perspective deviation as a \textit{local multiplicative adaptation} of the projection operator. Empirically, we found that human perspective is best modeled in a normalized world coordinate space, and not in image space. We therefore model human perspective at $\p$ as $\p \rightarrow \humanPerspD(\p) \mathbf{C} 
[\p;1]$, where a $\humanPerspD(\p)$ is a $4 \times 4$ deviation matrix, followed by perspective division by $p_w$. Input shapes are normalized so their origin is at $(0, 0, 0)$ and the shape is within the $[-1, 1]^{3}$ unit box. We parameterize  \humanPerspD using these normalized world coordinates. 

We represent \humanPerspD as an MLP that takes in 3D (normalized) coordinate information and outputs $15$ values. We reshape these values to a $4 \times 4$ matrix, with the last element set to $1$. Given this one-to-one relation, we use \humanPerspD to represent, based on context, both the human perspective matrix as well as the MLP output.

\paragraph*{Additional Deviation Properties} In addition to expecting our learned deviation functions \humanPerspD to change gradually across the input shapes,
we aim for them to preserve core properties of the projections of the depicted curves such as slope and shape; prior research on sketch analysis \cite{true2Form:2014,crossShade:12} suggests that artists seek to preserve these properties in their sketches.
Hence, while artists' local or global perspective deviation may be quite substantial, we do not require our deviation to be minimal across the board. 
Last but not least, we seek learned deviations that generalize across views and similar shapes; to this end, we explicitly seek deviations that are similar across similar views and are spatially smooth even when away from the input shape surface. 

\paragraph*{Algorithm Overview}
We learn a deviation matrix $\humanPerspD(\p)$, (illustrated in the inset)  from one or more pairs of a source sketch along with its corresponding 3D object and an estimated camera matrix that best aligns the sketch and camera views. 
We break the task into the following stages (\Cref{fig:overview}): 
(i)~matching between sketched strokes and  3D shape contour curves computed and analytically projected using the estimated camera; 
(ii)~modeling and learning a human perspective deviation function, as a spatially-conditioned MLP, that moves contour points toward their matched stroke positions; 
\begin{wrapfigure}[10]{l}{.325\columnwidth}
	\vspace*{-.1in}
	\includegraphics[width=.4\columnwidth]{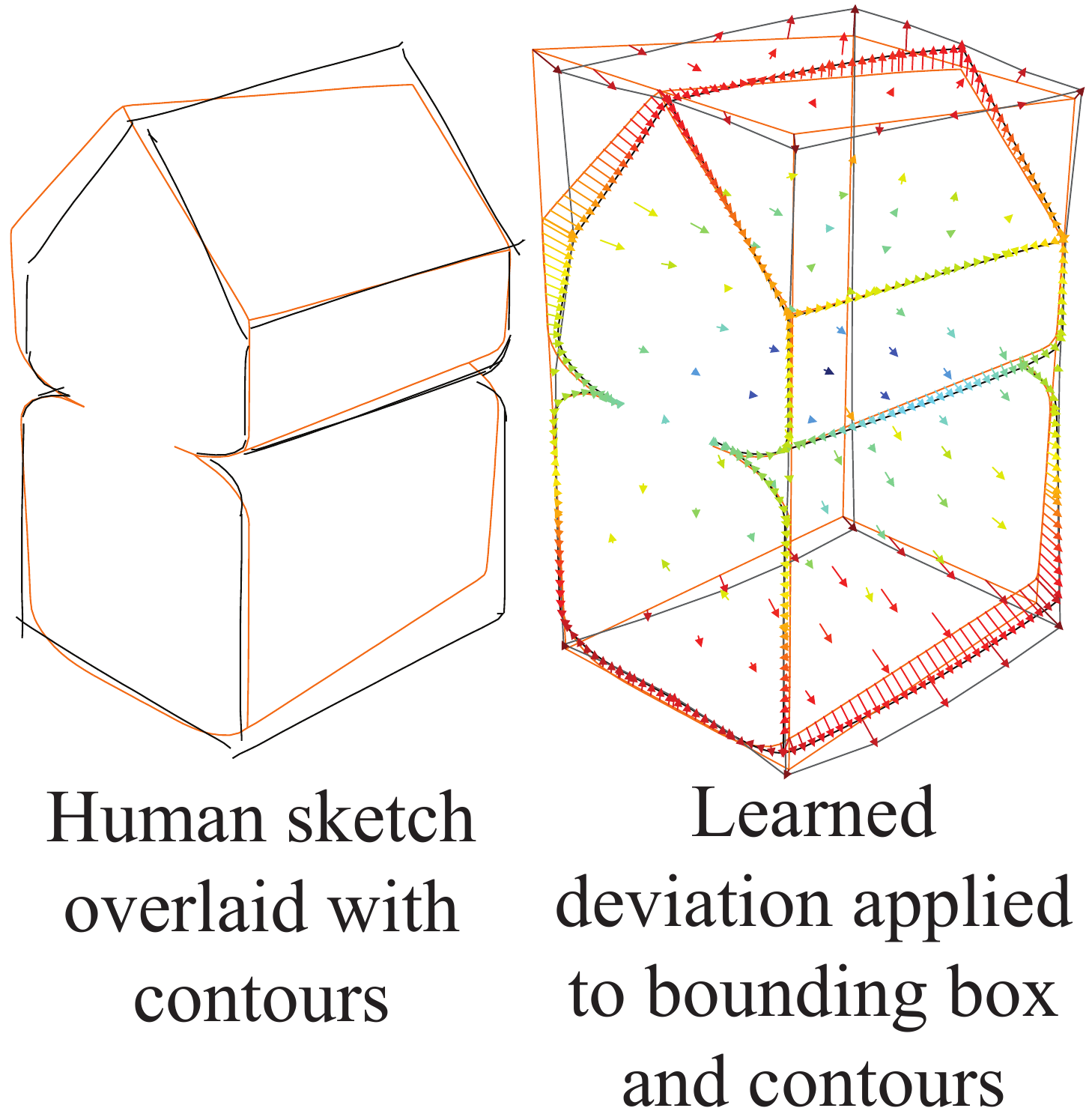}
\end{wrapfigure}
(iii)~regularizing the curves obtained by applying the learned perspective deviation, enforcing topological and other global constraints; and 
(iv)~self-augmentation to create additional data to retrain the MLP to better align $\humanPerspD(\p)$ with our global priors and better generalize to novel views. 
At inference, given a new shape and camera combination, we apply the learned perspective deviation function, followed by regularization (\Cref{fig:overview},right). 
We describe the method's stages below; see supplemental material for additional  details and hyper-parameter settings.

%% file: 4_algorithm.tex
\subsection{Preliminaries}
\label{sec:prelim}

We use the calibrated camera information to generate vector format occluding contours, sharp features, and surface boundaries of the input shape; referred to as {\em contours} throughout. We vectorized the input sketches, if needed, and resample both contour curves and sketch strokes using a fixed sampling interval rate (we denote the sampling interval length as $l$).  
In the following, unless stated otherwise, the term {\em curve} refers to a resampled projected contour curve and the term {\em stroke} refers to a resampled sketch stroke. 

We use $\hat p$ as the 3D locations of the 2D contour vertices $p$. 
Given a vertex $\hat p$ and a projection matrix $\humanPerspD$, the function $\proj(\hat p, \humanPerspD)$ is the 2D vertex computed by taking the matrix-vector product $\humanPerspD \mathbf{C}\cdot \hat p$ in homogenous coordinates, and applying the perspective divide.

\subsection{Matching Contour Curves to Sketch Strokes}
\label{subsec:matches}
First, we establish correspondences between vertices on the projected contour curves and vertices on the vectorized human sketch strokes. 
Intuitively, we seek to match contour vertices to nearby stroke vertices with similar tangents; we refer to this property as {\em compatibility}. 
One of the challenges in computing these correspondences is that they are not one-to-one (Fig~\ref{fig:overview}ab). Sketches may contain strokes, or portions of strokes, with no matching contours, due to oversketching~\cite{strokestrip:21}; contour curves may be depicted using multiple strokes, and some of the curve vertices may not have corresponding stroke locations. At the same time, while we do not expect exact one-to-one contour to stroke matches, we generally expect segments formed by consecutive contour vertices to match similarly consecutive stroke vertices, or in cases where a curve may correspond to multiple artist strokes or stroke sections, to match pairs of stroke vertices that form roughly parallel line segments; we refer to this property as {\em consistency}. 

\begin{figure}
    \centering
    \includegraphics[width=0.9\linewidth]{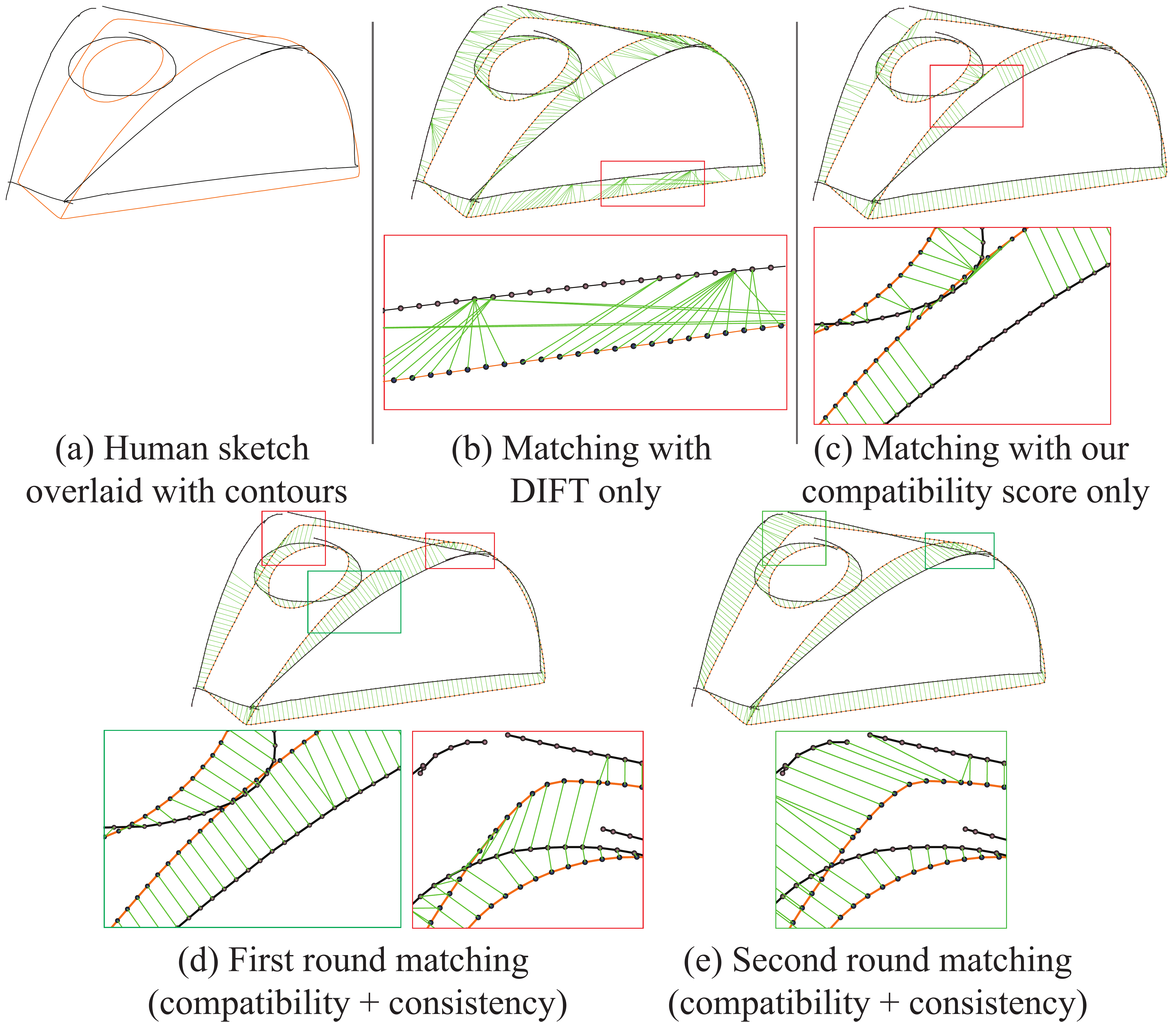}
\caption{We seek for artist-intended matches between contour curves and sketch strokes~(a). Using only vertex-to-vertex matching scores whether feature based (DIFT~\cite{tang2023emergent}) (b) or our compatibility based (c) produces locally optimal but globally poor matches; accounting for consistency (d) produces better matches at contour level  but can still lead to instances where multiple curves match the same stroke (see inset zoom). Our second matching step resolved these undesirable many-to-one matches~(e).}  
\label{fig:matching}
\vspace*{-.2in}
\end{figure}

This combination of requirements differs from the classical image space matching setting where users seek to match points with similar features  with no requirement for any consistency between the matches.
Dropping the consistency requirement makes the problem much simpler but results in matches that do not align with artist intent (\Cref{fig:matching}b uses DIFT ~\cite{tang2023emergent}, \Cref{fig:matching}c uses our compatibility score). 
We account for both compatibility and consistency by formulating matching as an instance of the classical Hidden Markov Model (HMM) problem \cite{matchingHMM:09} (\Cref{fig:matching}de). 
  
Given a contour curve $S := \{p_1,\dots p_n\}$, we first form, for each contour vertex $p_i$, a candidate set of potential matching stroke vertices $Q:=\{q_1,\dots q_m\}$ on the human sketch based on the distance between these vertices in 2D image space. We then evaluate potential matches $(p_i, q_ {j(i)})$ using a combined vertex-to-vertex \textit{compatibility} score $S^v(p_i, q_{j(i)})$ and a \textit{consistency} score $S^e(p_i, p_{i+1},q_{j(i)},q_{j(i+1)})$, which assesses compatibility between potential matches of consecutive curve vertices.  Using the classical HMM formulation, the overall score given by matching the vertices of $S$ to the vertices of $Q$ is:
\begin{equation}
M(S,Q):=\prod_{i}S^v(p_i, q_i)S^e(p_i, p_{i+1}, q_i, q_{i+1}).
\end{equation}
Using a product, rather than a sum, discourages outlier matches. 
We compute the matches for each curve that maximize $M(S,Q)$ using the Viterbi algorithm \shortcite{viterbi}. To obtain a valid solution, we exclude any vertices with empty matching candidate sets, and any edges emanating from such vertices, from the per-curve score. 

\textit{Compatibility Score ($S^v$).} Given a paired curve vertex $p$ and stroke vertex $q$, we define the score of using $q$ as the match of $p$ as a function of two terms, designed to be on the same scale, as:
\begin{eqnarray}
	d_a &=& \|p-q\|_{2} \nonumber \\
	d_t &=& 1 - | \mathcal{T}_S(p) \cdot  \mathcal{T}_Q(q)|.
\label{eq:2}
\end{eqnarray}
The first term is the absolute distance between them, while the  second term measures the similarity of the vertices' tangents and encourages matches that have  similar orientations:  here, $\mathcal{T}_{S}(p)$ is the normalized tangent vector of $S$ at $p$ (respectively for $Q$ at $q$). The overall score for matching $q \rightarrow p$ is thus:
\begin{equation}
	S^v(p, q) = e^{{-(d_a+d_t)^2}/{2 \sigma_1^2}}.
\label{eq:3}
\end{equation}

\textit{Consistency Score ($S^e$).}  We formulate consistency purely geometrically, and prioritize matching contour edges to pairs of vertices where the line connecting these vertices has similar length and orientation to the edge ones. Given a pair of consecutive vertices $p_i$ and $p_{i+1}$ (potentially) respectively matching a pair of vertices $q_{j(i)}$ and $q_{j(i+1)}$, we measure consistence $S^e(p_{i, i+1},q_i|q_{i+1})$ as:
\begin{eqnarray}
		d_p&=& \|(p_{i+1} - p_i) - (q_{j(i+1)} - q_{j(i)})\|_{2} \\ \nonumber
	       S^e(p_{i, i+1},q_i|q_{i+1}) &=& e^{{-d_p^2}/{2 \sigma_1^2}}. 
\end{eqnarray}

While we expect different contours to be depicted using different strokes, enforcing global matching constraints within the optimization framework above would dramatically increase algorithm complexity, making the matching problem NP-complete. 
Instead, we compute matches independently for each contour curve (Fig~\ref{fig:matching}d) and then identify and resolve cases where multiple curves or portions of curves are matched to the same stroke/stroke portion. We first identify vertices from different curves that match the same stroke vertex $q$ (\Cref{fig:matching}d,inset). We then do another matching round, where for these conflicted curve vertices we double the distance range to search in when finding candidate sets and exclude their previously matched stroke vertices from the candidate set. For each curve vertex in conflict, we then assign either its first round stroke match, or its new second round match, depending on which solution minimizes its score (\Cref{eq:3}), see \Cref{fig:matching}e.

\subsection{Learning Human Perspective}
\label{subsec:learning}
We use the matching results to learn a deviation function that approximately projects the original 3D locations $\hat p_i$ of the projected contour vertices $p_i$ to their matching 2D stroke vertices $q_{j(i)}$, while 
preserving contour slope, shape, and spatial smoothness. 

Formally, let $P:=\{p_1, \dots p_n\}$ be all vertices on the projected contours; let $Q:=\{q_1, \dots q_n\}$ be the matched vertices in the human vector sketch; for notational simplicity, we replace $q_{j(i)}$ with $q_i$. By abuse of notation, we identify each vertex $p_i$ with a distortion matrix $\mathbf{D_{i}}$ that is the output of the MLP at $\hat p_i$ (i.e., $\mathbf{D_{i}}=\mathbf{D(\hat p_i)}$), and denote the collection of all such matrices $\{\mathbf{D_1}, \dots \mathbf{D_n}\}$.  Our learned MLP takes as input the coordinate $\hat p$, in the normalized object space $[-1,1]^3$, and outputs a pointwise deviation matrix $\mathbf{D_{\hat p}}$.

\paragraph{Overall Loss} 
We learn an MLP that minimizes the following overall loss function, 
\begin{equation}
	L (D) := w_1 \cdot L_{data} + w_2 \cdot L_{shape}  +  w_3 \cdot L_{slope}  + w_4 \cdot L_S + w_5 \cdot L_{depth}.
\end{equation}
The first term aims to project 3D contour vertices close to their matching stroke vertices; the second and third terms preserve contour shape and slopes; the fourth term ensures deviation smoothness in 3D space; and the last, regularizer, term seeks to avoid depth instabilities by preserving relative point depth under our distortion function. We set $w_1=0.001,w_2= 10, w_3=w_4= 1, w_5=10^{-5}$.  This setting prioritises shape preservation above all other properties, and prioritizes our general priors about deviation above the data term. A small depth regularizer is sufficient to avoid instabilities.

\paragraph{Data Loss ($L_{data}$)} 

The term moves projected contour vertices $\proj(\hat p_i,\humanPerspD_i)$ toward their counterparts $q_i$ on the human sketch, and is defined as,
\begin{equation}
	L_{data} := \frac{1}{\textit{avg}_l}\frac{1}{n} \sum_{i \in n} \alpha_i \|\proj(\hat p_i,\humanPerspD_i) - q_i\|_{1}
\end{equation}
where $\textit{avg}_l = \frac{1}{n}\sum_{i \in n}{\| p_i - q_i\|_1} + \epsilon.$ 
We normalize each individual term by  a corresponding confidence value $\alpha_i$, and normalize the entire data term by the average distance between matching vertices (we add $\epsilon$ to avoid division by zero for perfect matches). 

\paragraph{Confidence.} While we  seek to project 3D contour vertices $\hat p_i$ toward their stroke matches $q_i$, the matches we compute may be imperfect due to factors such as oversketching (e.g., top of shampoo bottle in \Cref{fig:overview}b). We therefore associate confidence values $\alpha_i$ with each matched pair and use these to control the degree to which they are enforced. We base these values on the difference between intrinsic contour and stroke shape at the respective 2D vertices. Since artist sketches tend to preserve curve shape, mismatches between local contour/stroke curve shapes point to potential matching errors. We use polyline angles $\measuredangle (p_i)$ and $\measuredangle (q_i)$ at $p_i$ and $q_i$ respectively as proxy for shape (since we use uniform sampling these serve as an approximation of curvature). 
We define per-vertex confidence as:
\begin{equation}
\alpha_i = e^{{-(\measuredangle (p_i) - 
\measuredangle (q_i) ) ^2}/{2 \sigma_2 ^2}}.
\end{equation}

\paragraph{Shape Loss ($L_\textrm{shape}$)} The term aims to ensure our deviation preserves curve shape penalizing non-uniform scale and shear (see \cite{araujo22alup}), while allowing for uniform scale, as:
\begin{eqnarray}
    L_{\textrm{shape}}:= \sum_{i \in V} \sum_{j,k\in N(i);j\neq k} (1 - \alpha + \varepsilon) \|(\proj(\hat p_{k}, \humanPerspD_{k}) - \proj(\hat p_{i}, \humanPerspD_{i}) \nonumber \\
    - (\mathbf{R_i} \frac{\|p_{k} - p_i\|}{\|p_j- p_{i}\|} (\proj(\hat p_{j}, \humanPerspD_{j}) - \proj(\hat p_{i}, \humanPerspD_{i})))\|^2.
\end{eqnarray}
Here, $V$ is the set of all vertices lying on the interior of projected contour curves, and $N(i)$ are all neighbouring vertices of the vertex $i$; $\alpha$ is the minimum confidence value $\alpha_i$ of the three consecutive vertices $p_{j}, p_{i}, p_{k}$ forming the two edges; $R_i$ is the rotation matrix in 2D space that rotates the vector $p_j - p_{i}$ to $p_{k}-p_{i}$. We choose \textit{not} to penalize uniform scale as artists employ deviation that by design changes feature scale\cite{Hertzmann2024}.

\paragraph{Slope Loss ($L_{\textrm{slope}}$)}  The term aims to preserve the slopes of the projected contours under our learned deviation
\begin{equation}
    L_{\textrm{slope}} := \frac{1}{n-1} \sum_{i=2}^n \left( \hat{n}_{i}\cdot  \frac{(\proj(\hat p_{i}, \humanPerspD_{i}) - \proj(\hat p_{i-1}, \humanPerspD_{i-1}))}{\|p_i - p_{i-1}\|} \right)^2
\end{equation}
where $\hat{n}_{i}$ is the 2D normal of the projected contour edge ($p_i - p_{i-1}$). While artists do not strictly preserve slopes, the changes they introduce are typically subtle. 

\begin{figure*}
    \centering
    \includegraphics[width=.99\linewidth]{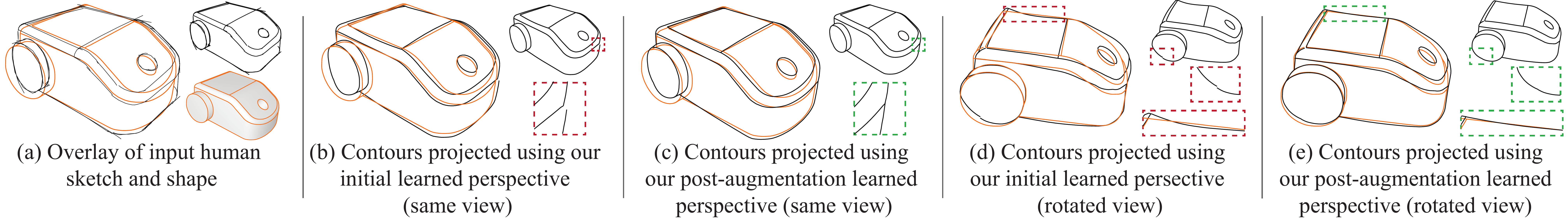}
    \caption{Self-augmentation impact: (a)~input sketch and analytically projected contours; (b)~same view inference output using first learned MLP; (c)~same view inference output using data augmented MLP,  alternative view~ inference output using first learned MLP~(d), alternative view inference output using data augmented MLP~(e). The impact of augmentation is more notable for further away camera views. }
    \label{fig:augmentation}
    \vspace{-.1in}
\end{figure*}

\paragraph{Smoothness Loss ($L_S$)} The term encourages smooth changes in the distortion matrix across view space and is computed over a set of points $\hat S$ containing all contour vertices $\hat p$ as well as the vertices of a dense grid spanning our 3D domain box $[-1,1]^3$. Given pairs if vertices $s_i \in \hat S$ and $s_j \in \hat S$, we express smoothness as the expectation for $\humanPerspD_i $ and $\humanPerspD_j $ to be increasingly similar for nearby vertices: 
\begin{equation}
L_S :=  \frac{1}{| \hat S|} \sum_{s_i \in\hat S} \sum_{s_j \in \hat S,i \neq j} e^{{-(\|s_i - s_j\|)^2}/{2 \sigma_1^2}} \|\humanPerspD_i - \humanPerspD_j\|_{\textrm{Frob}}.
\end{equation}

\paragraph{Depth Consistency Loss ($L_{\textrm{depth}}$)} In post-projection space, inverting depth axis direction, either globally or locally, has no impact on the 2D projection. Thus, such inversion is not penalized by any of the terms above. While unobservable for an individual view, inversion leads to inconsistent results when the camera is rotated. Our depth consistency loss  penalizes inversions by preserving the relative depth $\hat p_j^z$ of the transformed  vertices $\hat p_j$ along the viewspace z-axis:
\begin{equation}
L_{\textrm{depth}} :=  \sum_{i, j \in n, i \neq j} \old{w_{ij}} \| \new{(}D_i C \hat p_i^z - D_i C \hat p_j^z) - (\hat p_i^z - \hat p_j^z)\|.
\end{equation}

\subsection{Inference and Regularization}
\label{sec:infer_reg}
We use our learned MLP to render any set of 3D curves from a given camera view. Specifically, we use the MLP to project 3D surface vertices $\hat p$ to image space, by first multiplying each vertex by the user specified camera matrix $C$, multiplying the result by $D(\hat p)$ and then applying perspective divide (\Cref{fig:overview}c). Changing the perspective can change inter-contour occlusions and shift T-junction locations. We recover the correct contour topology via a post-inference regularization step (see supplemental material). 

\subsection{Self-Augmentation for Refined Learning}
Our initial learning 
is based on potentially imperfect contour-to-stroke matches and can bake in matching imperfections (e.g., top of the shampoo in \Cref{fig:overview}). It also indirectly promotes consistency across views, but this is not necessarily the case. To make learning more robust to matching errors and improve generalizability, we repeat the learning step using augmented data. \new{The primary goal of augmentation is to preserve the properties of the projected input contour curves that artists are known to preserve, thus obtaining output that appears human-like in all views.}

Specifically, we use our inference to generate new pairs of contours and matching deviated contours. We render the contours of the rotated shape twice, once using an analytical camera matrix $C_r$ and once using our inference method that multiples $C_r$ by our learned deviation matrix $\humanPerspD$, then regularizes the output. We then treat the deviated contours as the matched strokes of their analytical counterparts.  We then use these matched contour/''sketch'' pairs, plus the original contours and sketch, as training data for another learning step using the same MLP architecture and loss functions as the initial phase. \new{Our data term, by design, has a very small weight (0.001) relative to the weights of the shape, slope, and spatial smoothness terms (10,10,1). These terms balance fitting the input sketch against strong expectations that output contours preserve input contour shape and slopes, even from novel views. During augmentation, these terms act together to counteract data artifacts rather than propagating/reinforcing them, preventing us from overfitting to the input view matches.} 

In our experiment, we perform this step twice, first rotating the object by $[-5,-4,\ldots, 4, 5]$ degrees around the vertical axis and finetuning the MLP. We repeat this process, this time augmenting the data by rotating the object by $[-10,-9,\ldots,9,10]$. \new{We emperically chose the range [-10,10]: at [-5,5] we observed some artifacts under large rotations, while [-15,15] offered no additional benefit.} This iterative process enhances visual consistency across different views. \new{Without self-augmentation, small matching inaccuracies in the original view can trigger larger artifacts in close-by views. Removing augmentation produces wobbly unnatural looking curves, even for nearby views.}

%% file: 6_results.tex
\section{Results}
\label{sec:results}

\paragraph{Dataset and evaluation.}
We train our method on 169 individual pairs of sketches and corresponding shape contours: 96 from OpenSketch  \cite{OpenSketch19} (6 artists, 9 shapes, 1 or 2 views), 68 from \cite{xiao2022differsketching} (10 artists, 8 shapes), and 5 newly collected cube sketches from 5 artists. We then evaluate the resulting 169 models on both the input shape contours rendered from different views (e.g. \Cref{fig:resultsgallery}) and on contours computed on other shapes (e.g. \Cref{fig:cross_shape}). In addition to the representative examples shown in the paper, we include the outputs of models trained on all above pairs in the supplemental. In all figures we render contours projected via analytical perspective in  \textit{{\color{orange} orange}} and render artist sketches and contours projected using learned perspective (ours, other methods, and ablations) in \textit{black}. We overlay artist sketches and contours projected using learned perspective over the analytical contours, to visualize the deviation between them. See supplementary for details of all evaluations below. 

\paragraph{Alignment.}
Figs. \ref{fig:teaser}b,~\ref{fig:overview}e,~\ref{fig:augmentation}c,~\ref{fig:resultsgallery}b show our models learned on contour/sketch pairs applied to their training view contours. 
In each case, the overlays show that the deviation between contours projected using our method and the analytical ones is visually \textit{very similar} to that between the artist-sketched and analytical ones. We confirm this observation quantitatively: the $L_1$ chamfer distance between our contours and the artist-sketched ones is on average much smaller than that between the artist sketch and the analytical contours ($2.45e{-3}$ vs. $7.75e{-3}$, averaged over all models, all distances normalized by input image diagonal). This evaluation confirms that our inference results are indeed aligned with the training data. 

\paragraph{Generalization.}
We demonstrate that our method consistently generalizes to alternative views and unseen shapes in Figures \ref{fig:teaser}, \ref{fig:resultsgallery}~(views) and \ref{fig:teaser}, \ref{fig:cross_shape}, \ref{fig:same_view_diff_model}~(shapes). In all the figures, the overlays of our and analytical contours show visually similar deviation to that between the training sketches and their corresponding analytical contours. \Cref{fig:same_view_diff_model} demonstrates the outcome of applying different models to the same set of analytical contours; the resulting deviations are distinctly different and visually similar to those of their respective training pairs. 

\paragraph{Quantifying Consistency.} 
We quantitatively evaluate consistency across views as follows. We use our learned perspective deviation $D$ to render the contours of our input shape, with perspective deviation, from a new angle $\alpha$; we then learn a new perspective deviation $D'$ using these new contours as the input `sketch' and apply $D'$ to the original view contours (i.e. by rotating by $-\alpha$) and compute the Chamfer distance between these input contours rendered with D and those  rendered with D'; zero value indicating perfect consistency. Visually, the results are  very similar, especially for smaller $\alpha$, with the average Chamfer distances between paired contours ranging from 1.3e-1 for $\alpha=\pi/10$,  3.4e-3 for $\alpha=\pi/4$, and 3.8e-3 for $\alpha=\pi/2$. 

\paragraph{Perceptual Evaluation: Generalization}
We assess how well our method generalizes across shapes by examining if viewers can discern which model was used to generate a set of projected contours. 
Our study shows viewers an overlay of  a sketch and corresponding analytical contours on top, and two overlays of contours projected using learned perspective and their analytical counterparts below. One overlay shows our model trained on the input on top applied to the contours of an unseen shape; the second shows the result of applying a model trained on a different sketch/contour pair to the same contours.  Viewers were asked  ``Given the relationship between the  orange and black curves at the top which of the pairs of orange and black curves at the bottom exhibits a more similar relationship?''.  Viewers rate the outputs trained  on the input pairs on top as having more similar relation to the references 63\% of the time (21\% other, 2\% both, 14\% neither), demonstrating that ours generalizes deviation present in training sketches to other inputs.

\paragraph{Perceptual Evaluation: Human-Like Outputs.}  We also assessed if our contour outputs look more human like than those generated using analytic perspective. Study participants were shown same view contours projected using both analytical and our learned perspective, and were asked to assess which output was ``more likely to have been drawn by a human''.  Participants rated our outputs as more human-like 71\% of the time, the analytical contours as more  human like 12\% of the time, both equally human-like 8\%, neither human-like 9\%. The study confirms that our outputs look more human-like than analytical contours as desired.

\begin{figure*}[t!]
	\includegraphics[width=0.9\linewidth]{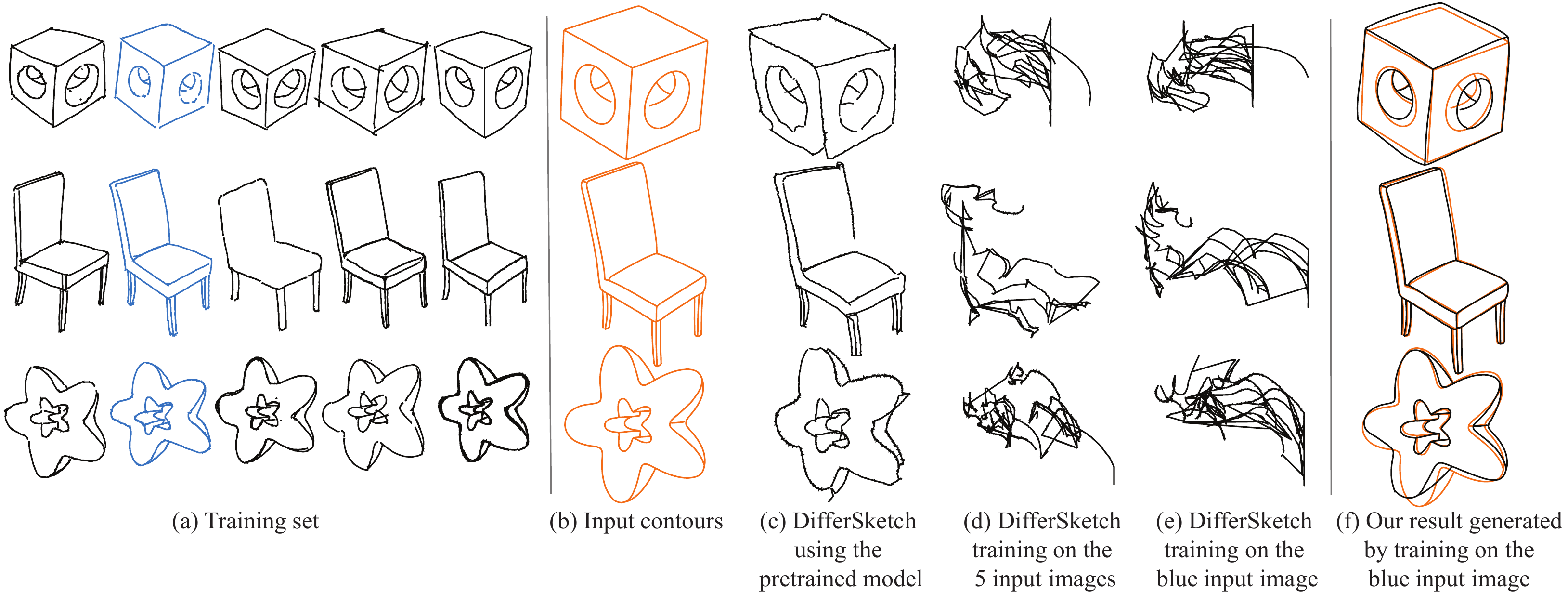}
\caption{Comparison versus DifferSketching \cite{xiao2022differsketching}. (a) same view/shape sketches in DifferSketching dataset; (b) analytically projected contours of the depicted shapes in a matching view; (c-e) Applying DifferSketching to the contours in (b): (c)~DifferSketching pre-trained on full data corpus; (d)~trained only on the sketches in (a) and the contours in (b); (e)~trained on the blue sketch in (a) and the contour in (b); (f) Our outputs trained the blue sketch in (a) and the contour in (b). In all cases, DifferSketching outputs are distorted, and quality reduces as training set size decreases.  Our method capture the artists' deviation and generalizes.} 
\label{fig:differ}
\vspace*{-.15in}
\end{figure*}

\paragraph{Comparison to Prior Work} 
\Cref{fig:differ} compares our results to those of DifferSketching~\cite{xiao2022differsketching} on input sketch and contour pairs that are part of their training dataset. \new{In \Cref{fig:differ}c, we show the output of DifferSketching's pretrained model.} \old{The DifferSketching outputs (using their model} \new{DifferSketching's pretrained model (}trained on their entire dataset) cannot capture individual artist choices, exhibit high-frequency noise, and deviate significantly from the corresponding artist sketches (\Cref{fig:differ}c). Retraining their model on a subset of the dataset (all sketches of the input shape (\Cref{fig:differ}d), or a single sketch/contour pair (\Cref{fig:differ}e)) \new{using their default parameters (extrinsic noise = intrinsic noise = 0.3)} dramatically reduces result quality. \new{We also tried using smaller values for extrinsic and intrinsic noise (both = 0.001), and setting each of the two parameters to 0 and the other to 0.001 or 0.3. Using their model with the above settings brings the output contours spatially closer to the input when compared to the default but retains undesirable stroke "jaggies" and other stroke-level artifacts. In all cases, learning on a single input or single family of inputs (e.g. all drawings of the same shape) produces catastrophic failures like those shown in \Cref{fig:differ}de. Training DifferSketch by fine-tuning their pretrained model on either one input, or a small family of inputs, produces similar catastrophic failures as \Cref{fig:differ}de.} Our method (\Cref{fig:differ}f) successfully trains on their inputs and generalizes across shapes.  We conclude that while DifferSketching can be beneficial for other applications, it is unsuitable for ours.

For completeness, in the supplementary, we provide additional visual comparisons to methods for novel view synthesis \cite{liu2023zero1to3}, generating stylized line drawings from images \cite{chan2022drawings}, and to Pix2Pix~\cite{pix2pix2017} trained on our paired sketch and contour corpus. In all cases, our experiments show that these alternative methods fail to reproduce human perspective deviation, showcasing both the need for a method that explicitly learns perspective deviation, instead of stylization, and ours ability to do so.

\paragraph{Ablations.} 
We compared our method against a 2D version of our MLP, which attempts to learn pure 2D deviation between artist and analytical contours. While the outputs generated using these 2D only models appear reasonable in the input and nearby views, the models do not generalize to farther away views where their outputs exhibit undesirable deformation (Fig~\ref{fig:2dmlp}). We validated this observation via two studies. 
The first study asked participants whether our outputs, or ones generated using the 2D MLP, were more accurate rotated depictions of the original sketched shape. Participants ranked ours as more accurate 79\% of the time (vs 6\% 2D MLP, 7\% both, 8\% neither).  The second study used a similar design to the generalization study above, but compared our output to that of the 2D MLP.  Participants assessed the relations between our output and analytical contours as more similar to the reference than the one between 2D MLP outputs and analytical contours 72\% of the time versus 17\% for 2D MLP (6\% both, 5\% neither). These studies confirm the need for a 3D approach for encoding and learning perspective deviation. 

\Cref{fig:training_across_shapes} evaluates our decision to use single sketch-contour pairs to learn meaningful deviation function by training our perspective MLP on larger data corpuses.  As demonstrated, increasing the size of the training corpus diminishes the magnitude of the deviation; contrary to our goal of capturing training sketch deviation. See supplemental for details and additional ablations.

\paragraph{Applications}
\new{Learning human perspective can support a range of downstream applications. One natural direction is non-photorealistic rendering, where our approach can add a human touch to generated outputs in combination with other stylization elements. Another is content addition or editing (for instance, through generative AI), where the additions/edits would obey and respect the perspective deviations present in the input, ensuring visually coherent results. Our framework could also serve as a tool for analyzing artistic choices, such as whether artists apply similar perspective deviations across views or shapes.}

%% file: 7_conclusion.tex
\section{Conclusion}
\label{sec:conclusion}

We  presented the first method to model and learn the perspective projection humans use when creating line drawings. As demonstrated, our method faithfully captures input sketch artist perspective and generalizes it across novel views and similar shapes. 

\paragraph{Limitations and Future Work} 
The only notable failure scenarios we observed were either due to faulty vector contour extraction from our input shapes or poor input camera estimation for the input shape/sketches.
Our method produces a single learned perspective deviation per input, leaving it to the user to pick the input they like best. It would be interesting to analyse the similarities and differences between the deviation functions we learn across different shapes and artists and to use the results of this analysis to select or compute the visually best deviation for a new input.
 
An exciting follow-up avenue for our work is to use our approach for learning deviations  as a basis for a method capable of inverting artist deviation, i.e., taking free-hand sketches and producing 2D curves that are the {\em analytic} projection of artist intended 3D surface curves. Being able to invert deviation will improve the robustness of  sketch-based 3D modeling.

%% file: acks.tex
\begin{acks}
	We thank Congyi Zhang for their help with figures. We acknowledge the support of the Natural Sciences and Engineering Research Council of Canada (NSERC) grant RGPIN-2024-03981. Finally, this work is supported in part by the Institute for Computing, Information and Cognitive Systems (ICICS) and Advanced Research Computing (ARC) at UBC.
\end{acks}

%% file: figures.tex
\begin{figure*}[t]
    \centering
    \includegraphics[width=.8\linewidth]{figures_comp/results_v11.pdf}
    \caption{A gallery of our results. We show (a) overlay of human sketch\new{es} and shape contours in best matching view (insets show the sketch\new{es} and contours separately); (b) our learned output\new{s} under the same view as (a); (c) Applying our learned perspective\new{s} to contours in different other rotated views.  In all examples our outputs match the sketch\old{'s}\new{es'} perspective deviation\new{s}.
    }
    \label{fig:resultsgallery}
\end{figure*}

\begin{figure*}
\centering
\includegraphics[width=.92\linewidth]{figures_comp/2D_MLP_v3.pdf}
\caption{Ablation against 2D MLP: (a,\old{c}\new{d}) input sketch\new{es} and matching view contours; (b,\old{d}\new{e})  rotated view outputs generated using 2D MLP; (c,\old{e}\new{f}) our results for same views. 2D MLP outputs undesirably and notably diverge from the analytical contours in such views. Our results maintain the  same deviation\new{s} relative to analytical contours, as the input sketch\new{es} across all views. } 
\label{fig:2dmlp}
\end{figure*}

\begin{figure*}[t]
	\centering
	\includegraphics[width=.98\linewidth]{figures_comp/training_across_shapes_and_artists_v4.pdf}
	\caption{Learning from multiple sketch/shape pairs:
	 Left: learning from 18 sketches by the same artist: (a) input\new{s}, (b) our output\new{s}, (c) output\new{s} of learning from 18 sketches.
	 The result captures the common characteristic\old{'}s of artist\new{'}s perspective, but is more subtle then one learned form a single input pair.   Right: learning from same shape, approximately same view sketches from 6 artists: (d) input, (e) our output, (f) output of learning from 6 sketches. The result l\old{o}oses the expressiveness, with learned perspective essentially identical to analytical one.}
	\label{fig:training_across_shapes}
	\label{fig:training_across_artists}
\end{figure*}

\begin{figure*}[t]
	\centering
	\includegraphics[width=.99\linewidth]{figures_comp/cross_shape_results_v2.pdf}
	\caption{Applying learned perspective to new shapes:  (a,c,e) overlay of human sketch\new{es} and shape contours in best matching view ; (b,d,f) Applying our learned perspective\new{s} to contours of other shapes in different views.}
	\label{fig:cross_shape}
\end{figure*}

\begin{figure*}[t]
	\centering
	\includegraphics[width=.99\linewidth]{figures_comp/same_view_different_model_v2.pdf}
	\caption{Applying different learned perspectives to the same shape. As desired, applying different learned perspectives to the same analytical contours produces distinctly different projections, each consistent with its source training sketch projection.}
	\label{fig:same_view_diff_model}
\end{figure*}

%% file: 0_main.bbl

\begin{thebibliography}{56}


\ifx \showCODEN    \undefined \def \showCODEN     #1{\unskip}     \fi
\ifx \showDOI      \undefined \def \showDOI       #1{#1}\fi
\ifx \showISBNx    \undefined \def \showISBNx     #1{\unskip}     \fi
\ifx \showISBNxiii \undefined \def \showISBNxiii  #1{\unskip}     \fi
\ifx \showISSN     \undefined \def \showISSN      #1{\unskip}     \fi
\ifx \showLCCN     \undefined \def \showLCCN      #1{\unskip}     \fi
\ifx \shownote     \undefined \def \shownote      #1{#1}          \fi
\ifx \showarticletitle \undefined \def \showarticletitle #1{#1}   \fi
\ifx \showURL      \undefined \def \showURL       {\relax}        \fi
\providecommand\bibfield[2]{#2}
\providecommand\bibinfo[2]{#2}
\providecommand\natexlab[1]{#1}
\providecommand\showeprint[2][]{arXiv:#2}

\bibitem[Ara\'{u}jo et~al\mbox{.}(2022)]%
        {araujo22alup}
\bibfield{author}{\bibinfo{person}{Chrystiano Ara\'{u}jo},
  \bibinfo{person}{Nicholas Vining}, \bibinfo{person}{Enrique Rosales},
  \bibinfo{person}{Giorgio Gori}, {and} \bibinfo{person}{Alla Sheffer}.}
  \bibinfo{year}{2022}\natexlab{}.
\newblock \showarticletitle{As-Locally-Uniform-As-Possible Reshaping of Vector
  Clip-Art}.
\newblock \bibinfo{journal}{\emph{ACM Transaction on Graphics}}
  \bibinfo{volume}{41}, \bibinfo{number}{4} (\bibinfo{year}{2022}).
\newblock
\urldef\tempurl%
\url{https://doi.org/0.1145/3528223.3530098}
\showDOI{\tempurl}


\bibitem[Bae et~al\mbox{.}(2008)]%
        {bae2008ilovesketch}
\bibfield{author}{\bibinfo{person}{Seok-Hyung Bae}, \bibinfo{person}{Ravin
  Balakrishnan}, {and} \bibinfo{person}{Karan Singh}.}
  \bibinfo{year}{2008}\natexlab{}.
\newblock \showarticletitle{ILoveSketch: as-natural-as-possible sketching
  system for creating 3d curve models}. In
  \bibinfo{booktitle}{\emph{Proceedings of the 21st annual ACM symposium on
  User interface software and technology}}. \bibinfo{pages}{151--160}.
\newblock


\bibitem[Bessmeltsev and Liu(2024)]%
        {BessmeltsevLiuSGP2024}
\bibfield{author}{\bibinfo{person}{Mikhail Bessmeltsev} {and}
  \bibinfo{person}{Chenxi Liu}.} \bibinfo{year}{2024}\natexlab{}.
\newblock \showarticletitle{Fundamentals and Applications of Sketch
  Processing}. In \bibinfo{booktitle}{\emph{Symposium on Graphics Processing
  (SGP 2024) Course}}.
\newblock
\urldef\tempurl%
\url{https://school.geometryprocessing.org/summerschool-2024/}
\showURL{%
\tempurl}


\bibitem[Chan et~al\mbox{.}(2022)]%
        {chan2022drawings}
\bibfield{author}{\bibinfo{person}{Caroline Chan}, \bibinfo{person}{Frédo
  Durand}, {and} \bibinfo{person}{Phillip Isola}.}
  \bibinfo{year}{2022}\natexlab{}.
\newblock \showarticletitle{Learning to generate line drawings that convey
  geometry and semantics}. In \bibinfo{booktitle}{\emph{CVPR}}.
\newblock


\bibitem[Choi et~al\mbox{.}(2024)]%
        {Choi_2024}
\bibfield{author}{\bibinfo{person}{Changwoon Choi}, \bibinfo{person}{Jaeah
  Lee}, \bibinfo{person}{Jaesik Park}, {and} \bibinfo{person}{Young~Min Kim}.}
  \bibinfo{year}{2024}\natexlab{}.
\newblock \showarticletitle{3Doodle: Compact Abstraction of Objects with 3D
  Strokes}.
\newblock \bibinfo{journal}{\emph{{ACM TOG}}} \bibinfo{volume}{43},
  \bibinfo{number}{4} (\bibinfo{date}{July} \bibinfo{year}{2024}),
  \bibinfo{pages}{1–13}.
\newblock


\bibitem[Cole et~al\mbox{.}(2008)]%
        {Cole:2008:WDP}
\bibfield{author}{\bibinfo{person}{Forrester Cole}, \bibinfo{person}{Aleksey
  Golovinskiy}, \bibinfo{person}{Alex Limpaecher},
  \bibinfo{person}{Heather~Stoddart Barros}, \bibinfo{person}{Adam
  Finkelstein}, \bibinfo{person}{Thomas Funkhouser}, {and}
  \bibinfo{person}{Szymon Rusinkiewicz}.} \bibinfo{year}{2008}\natexlab{}.
\newblock \showarticletitle{Where Do People Draw Lines?}
\newblock \bibinfo{journal}{\emph{ACM Transactions on Graphics (Proc.
  SIGGRAPH)}} \bibinfo{volume}{27}, \bibinfo{number}{3} (\bibinfo{date}{Aug.}
  \bibinfo{year}{2008}).
\newblock


\bibitem[Coleman et~al\mbox{.}(2005)]%
        {patrick:perspective:05}
\bibfield{author}{\bibinfo{person}{Patrick Coleman}, \bibinfo{person}{Karan
  Singh}, \bibinfo{person}{Leon Barrett}, \bibinfo{person}{Nisha Sudarsanam},
  {and} \bibinfo{person}{Cindy Grimm}.} \bibinfo{year}{2005}\natexlab{}.
\newblock \showarticletitle{3D screen-space widgets for non-linear projection}.
  In \bibinfo{booktitle}{\emph{Proc. Computer Graphics and Interactive
  Techniques}} \emph{(\bibinfo{series}{GRAPHITE '05})}.
  \bibinfo{pages}{221–228}.
\newblock


\bibitem[De~Paoli and Singh(2015)]%
        {de2015secondskin}
\bibfield{author}{\bibinfo{person}{Chris De~Paoli} {and} \bibinfo{person}{Karan
  Singh}.} \bibinfo{year}{2015}\natexlab{}.
\newblock \showarticletitle{SecondSkin: sketch-based construction of layered 3D
  models}.
\newblock \bibinfo{journal}{\emph{ACM Transactions on Graphics (TOG)}}
  \bibinfo{volume}{34}, \bibinfo{number}{4} (\bibinfo{year}{2015}),
  \bibinfo{pages}{1--10}.
\newblock


\bibitem[DeCarlo et~al\mbox{.}(2003)]%
        {DeCarlo:2003:SCF}
\bibfield{author}{\bibinfo{person}{Doug DeCarlo}, \bibinfo{person}{Adam
  Finkelstein}, \bibinfo{person}{Szymon Rusinkiewicz}, {and}
  \bibinfo{person}{AnthonySantella}.} \bibinfo{year}{2003}\natexlab{}.
\newblock \showarticletitle{Suggestive Contours for Conveying Shape}.
\newblock \bibinfo{journal}{\emph{Proc. SIGGRAPH}} \bibinfo{volume}{22},
  \bibinfo{number}{3} (\bibinfo{date}{jul} \bibinfo{year}{2003}),
  \bibinfo{pages}{848--855}.
\newblock


\bibitem[Dvoro\v{z}\v{n}\'{a}k et~al\mbox{.}(2020)]%
        {MonsterMash:2020}
\bibfield{author}{\bibinfo{person}{Marek Dvoro\v{z}\v{n}\'{a}k},
  \bibinfo{person}{Daniel S\'{y}kora}, \bibinfo{person}{Cassidy Curtis},
  \bibinfo{person}{Brian Curless}, \bibinfo{person}{Olga Sorkine-Hornung},
  {and} \bibinfo{person}{David Salesin}.} \bibinfo{year}{2020}\natexlab{}.
\newblock \showarticletitle{{Monster Mash}: {A} Single-View Approach to Casual
  3{D} Modeling and Animation}.
\newblock \bibinfo{journal}{\emph{Proc. of SIGGRAPH ASIA}}
  \bibinfo{volume}{39}, \bibinfo{number}{6}, Article \bibinfo{articleno}{214}
  (\bibinfo{year}{2020}).
\newblock


\bibitem[Eissen and Steur(2008)]%
        {eissen08}
\bibfield{author}{\bibinfo{person}{Koos Eissen} {and} \bibinfo{person}{Roselien
  Steur}.} \bibinfo{year}{2008}\natexlab{}.
\newblock \bibinfo{booktitle}{\emph{Sketching: Drawing Techniques for Product
  Designers}}.
\newblock \bibinfo{publisher}{Bis Publishers}.
\newblock


\bibitem[Eissen and Steur(2011)]%
        {eissen2011sketching}
\bibfield{author}{\bibinfo{person}{Koos Eissen} {and} \bibinfo{person}{Roselien
  Steur}.} \bibinfo{year}{2011}\natexlab{}.
\newblock \bibinfo{booktitle}{\emph{Sketching: The Basics}}.
\newblock \bibinfo{publisher}{Bis Publishers}.
\newblock


\bibitem[Foley et~al\mbox{.}(1996)]%
        {foley1996computer}
\bibfield{author}{\bibinfo{person}{James~D. Foley}, \bibinfo{person}{Andries
  van Dam}, \bibinfo{person}{Steven~K. Feiner}, {and} \bibinfo{person}{John~F.
  Hughes}.} \bibinfo{year}{1996}\natexlab{}.
\newblock \bibinfo{booktitle}{\emph{Computer Graphics: Principles and Practice}
  (\bibinfo{edition}{2nd} ed.)}.
\newblock \bibinfo{publisher}{Addison-Wesley}, \bibinfo{address}{Reading, MA}.
\newblock
\showISBNx{978-0201848403}


\bibitem[Gatys et~al\mbox{.}(2016)]%
        {StyleTransfer7780634}
\bibfield{author}{\bibinfo{person}{Leon~A. Gatys},
  \bibinfo{person}{Alexander~S. Ecker}, {and} \bibinfo{person}{Matthias
  Bethge}.} \bibinfo{year}{2016}\natexlab{}.
\newblock \showarticletitle{Image Style Transfer Using Convolutional Neural
  Networks}. In \bibinfo{booktitle}{\emph{2016 IEEE Conference on Computer
  Vision and Pattern Recognition (CVPR)}}. \bibinfo{pages}{2414--2423}.
\newblock
\urldef\tempurl%
\url{https://doi.org/10.1109/CVPR.2016.265}
\showDOI{\tempurl}


\bibitem[Gombrich(1951)]%
        {Gombrich1951-GOMTSO-3}
\bibfield{author}{\bibinfo{person}{E.~H. Gombrich}.}
  \bibinfo{year}{1951}\natexlab{}.
\newblock \showarticletitle{The Story of Art}.
\newblock \bibinfo{journal}{\emph{Journal of Aesthetics and Art Criticism}}
  \bibinfo{volume}{9}, \bibinfo{number}{4} (\bibinfo{year}{1951}),
  \bibinfo{pages}{339--340}.
\newblock
\urldef\tempurl%
\url{https://doi.org/10.2307/426517}
\showDOI{\tempurl}


\bibitem[Gryaditskaya et~al\mbox{.}(2020)]%
        {Lift3DSA20}
\bibfield{author}{\bibinfo{person}{Yulia Gryaditskaya}, \bibinfo{person}{Felix
  H\"{a}hnlein}, \bibinfo{person}{Chenxi Liu}, \bibinfo{person}{Alla Sheffer},
  {and} \bibinfo{person}{Adrien Bousseau}.} \bibinfo{year}{2020}\natexlab{}.
\newblock \showarticletitle{Lifting Freehand Concept Sketches into 3D}.
\newblock \bibinfo{journal}{\emph{ACM Transactions on Graphics (Proc. SIGGRAPH
  Asia)}} (\bibinfo{year}{2020}).
\newblock


\bibitem[Gryaditskaya et~al\mbox{.}(2019)]%
        {OpenSketch19}
\bibfield{author}{\bibinfo{person}{Yulia Gryaditskaya}, \bibinfo{person}{Mark
  Sypesteyn}, \bibinfo{person}{Jan~Willem Hoftijzer}, \bibinfo{person}{Sylvia
  Pont}, \bibinfo{person}{Fr\'{e}do Durand}, {and} \bibinfo{person}{Adrien
  Bousseau}.} \bibinfo{year}{2019}\natexlab{}.
\newblock \showarticletitle{OpenSketch: A Richly-Annotated Dataset of Product
  Design Sketches}.
\newblock \bibinfo{journal}{\emph{ACM Transactions on Graphics (Proc. SIGGRAPH
  Asia)}}  \bibinfo{volume}{38} (\bibinfo{date}{11} \bibinfo{year}{2019}).
\newblock


\bibitem[H{\"a}hnlein et~al\mbox{.}(2022)]%
        {hahnlein2022symmetry}
\bibfield{author}{\bibinfo{person}{Felix H{\"a}hnlein}, \bibinfo{person}{Yulia
  Gryaditskaya}, \bibinfo{person}{Alla Sheffer}, {and} \bibinfo{person}{Adrien
  Bousseau}.} \bibinfo{year}{2022}\natexlab{}.
\newblock \showarticletitle{Symmetry-driven 3D reconstruction from concept
  sketches}. In \bibinfo{booktitle}{\emph{ACM SIGGRAPH 2022 Conference
  Proceedings}}. \bibinfo{pages}{1--8}.
\newblock


\bibitem[H\"{a}hnlein et~al\mbox{.}(2022)]%
        {CAD2Sketch10.1145/3550454.3555488}
\bibfield{author}{\bibinfo{person}{Felix H\"{a}hnlein},
  \bibinfo{person}{Changjian Li}, \bibinfo{person}{Niloy~J. Mitra}, {and}
  \bibinfo{person}{Adrien Bousseau}.} \bibinfo{year}{2022}\natexlab{}.
\newblock \showarticletitle{CAD2Sketch: Generating Concept Sketches from CAD
  Sequences}.
\newblock \bibinfo{journal}{\emph{ACM Trans. Graph.}} \bibinfo{volume}{41},
  \bibinfo{number}{6}, Article \bibinfo{articleno}{279} (\bibinfo{date}{Nov.}
  \bibinfo{year}{2022}), \bibinfo{numpages}{18}~pages.
\newblock
\showISSN{0730-0301}
\urldef\tempurl%
\url{https://doi.org/10.1145/3550454.3555488}
\showDOI{\tempurl}


\bibitem[Hennessey et~al\mbox{.}(2017)]%
        {hennessey:sketch:17}
\bibfield{author}{\bibinfo{person}{James~W. Hennessey}, \bibinfo{person}{Han
  Liu}, \bibinfo{person}{Holger Winnemöller}, \bibinfo{person}{Mira
  Dontcheva}, {and} \bibinfo{person}{Niloy~J. Mitra}.}
  \bibinfo{year}{2017}\natexlab{}.
\newblock \showarticletitle{How2Sketch: Generating Easy-To-Follow Tutorials for
  Sketching 3D Objects}.
\newblock \bibinfo{journal}{\emph{Symposium on Interactive 3D Graphics and
  Games}} (\bibinfo{year}{2017}).
\newblock


\bibitem[Hertz et~al\mbox{.}(2023)]%
        {hertz2023StyleAligned}
\bibfield{author}{\bibinfo{person}{Amir Hertz}, \bibinfo{person}{Andrey
  Voynov}, \bibinfo{person}{Shlomi Fruchter}, {and} \bibinfo{person}{Daniel
  Cohen-Or}.} \bibinfo{year}{2023}\natexlab{}.
\newblock \showarticletitle{Style Aligned Image Generation via Shared
  Attention}.
\newblock  (\bibinfo{year}{2023}).
\newblock


\bibitem[Hertzmann(2024)]%
        {Hertzmann2024}
\bibfield{author}{\bibinfo{person}{Aaron Hertzmann}.}
  \bibinfo{year}{2024}\natexlab{}.
\newblock \showarticletitle{Toward a theory of perspective perception in
  pictures}.
\newblock \bibinfo{journal}{\emph{Journal of Vision}} \bibinfo{volume}{24},
  \bibinfo{number}{4} (\bibinfo{date}{04} \bibinfo{year}{2024}),
  \bibinfo{pages}{23--23}.
\newblock


\bibitem[Igarashi et~al\mbox{.}(1999)]%
        {Teddy10.1145/311535.311602}
\bibfield{author}{\bibinfo{person}{Takeo Igarashi}, \bibinfo{person}{Satoshi
  Matsuoka}, {and} \bibinfo{person}{Hidehiko Tanaka}.}
  \bibinfo{year}{1999}\natexlab{}.
\newblock \showarticletitle{Teddy: a sketching interface for 3D freeform
  design}. In \bibinfo{booktitle}{\emph{Proceedings of the 26th Annual
  Conference on Computer Graphics and Interactive Techniques}}
  \emph{(\bibinfo{series}{SIGGRAPH '99})}. \bibinfo{publisher}{ACM
  Press/Addison-Wesley Publishing Co.}, \bibinfo{address}{USA},
  \bibinfo{pages}{409–416}.
\newblock
\showISBNx{0201485605}


\bibitem[Isola et~al\mbox{.}(2017)]%
        {pix2pix2017}
\bibfield{author}{\bibinfo{person}{Phillip Isola}, \bibinfo{person}{Jun-Yan
  Zhu}, \bibinfo{person}{Tinghui Zhou}, {and} \bibinfo{person}{Alexei~A
  Efros}.} \bibinfo{year}{2017}\natexlab{}.
\newblock \showarticletitle{Image-to-Image Translation with Conditional
  Adversarial Networks}.
\newblock \bibinfo{journal}{\emph{CVPR}} (\bibinfo{year}{2017}).
\newblock


\bibitem[Kara and Shimada(2007)]%
        {kara2007sketch}
\bibfield{author}{\bibinfo{person}{Levent~Burak Kara} {and}
  \bibinfo{person}{Kenji Shimada}.} \bibinfo{year}{2007}\natexlab{}.
\newblock \showarticletitle{Sketch-based 3D-shape creation for industrial
  styling design}.
\newblock \bibinfo{journal}{\emph{IEEE Computer Graphics and Applications}}
  \bibinfo{volume}{27}, \bibinfo{number}{1} (\bibinfo{year}{2007}),
  \bibinfo{pages}{60--71}.
\newblock


\bibitem[Kemp(1991)]%
        {warwick22265}
\bibfield{author}{\bibinfo{person}{M. Kemp}.} \bibinfo{year}{1991}\natexlab{}.
\newblock \bibinfo{title}{The Science of Art -- Optical Themes in Western Art
  from Brunelleschi to Seurat}.
\newblock , \bibinfo{numpages}{617--619}~pages.
\newblock


\bibitem[Koenderink et~al\mbox{.}(2016)]%
        {Koenderink2016}
\bibfield{author}{\bibinfo{person}{Jan Koenderink}, \bibinfo{person}{Andrea van
  Doorn}, \bibinfo{person}{Baingio Pinna}, {and} \bibinfo{person}{Robert
  Pepperell}.} \bibinfo{year}{2016}\natexlab{}.
\newblock \showarticletitle{{On right and wrong drawings}}.
\newblock \bibinfo{journal}{\emph{Art and Perception}} (\bibinfo{date}{9}
  \bibinfo{year}{2016}).
\newblock


\bibitem[Koenderink and van Doorn(1991)]%
        {Koenderink:91}
\bibfield{author}{\bibinfo{person}{Jan~J. Koenderink} {and}
  \bibinfo{person}{Andrea~J. van Doorn}.} \bibinfo{year}{1991}\natexlab{}.
\newblock \showarticletitle{Affine structure from motion}.
\newblock \bibinfo{journal}{\emph{J. Opt. Soc. Am. A}} \bibinfo{volume}{8},
  \bibinfo{number}{2} (\bibinfo{date}{Feb} \bibinfo{year}{1991}),
  \bibinfo{pages}{377--385}.
\newblock


\bibitem[Kubovy(1986)]%
        {kubovy1986psychology}
\bibfield{author}{\bibinfo{person}{Michael Kubovy}.}
  \bibinfo{year}{1986}\natexlab{}.
\newblock \bibinfo{booktitle}{\emph{The Psychology of Perspective and
  Renaissance Art}}.
\newblock \bibinfo{publisher}{Cambridge University Press},
  \bibinfo{address}{Cambridge, UK}.
\newblock
\showISBNx{9780521302302}


\bibitem[Li et~al\mbox{.}(2022)]%
        {li2022free2cad}
\bibfield{author}{\bibinfo{person}{Changjian Li}, \bibinfo{person}{Hao Pan},
  \bibinfo{person}{Adrien Bousseau}, {and} \bibinfo{person}{Niloy~J Mitra}.}
  \bibinfo{year}{2022}\natexlab{}.
\newblock \showarticletitle{Free2cad: Parsing freehand drawings into cad
  commands}.
\newblock \bibinfo{journal}{\emph{ACM (TOG)}} \bibinfo{volume}{41},
  \bibinfo{number}{4} (\bibinfo{year}{2022}), \bibinfo{pages}{1--16}.
\newblock


\bibitem[Li et~al\mbox{.}(2018)]%
        {Li:2018:SketchCNN}
\bibfield{author}{\bibinfo{person}{Changjian Li}, \bibinfo{person}{Hao Pan},
  \bibinfo{person}{Yang Liu}, \bibinfo{person}{Alla Sheffer}, {and}
  \bibinfo{person}{Wenping Wang}.} \bibinfo{year}{2018}\natexlab{}.
\newblock \showarticletitle{Robust Flow-Guided Neural Prediction for
  Sketch-Based Freeform Surface Modeling}.
\newblock \bibinfo{journal}{\emph{ACM Trans. Graph. (SIGGRAPH ASIA)}}
  \bibinfo{volume}{37}, \bibinfo{number}{6} (\bibinfo{year}{2018}),
  \bibinfo{pages}{238:1--238:12}.
\newblock
\urldef\tempurl%
\url{https://doi.org/10.1145/3272127.3275051}
\showDOI{\tempurl}


\bibitem[Liao et~al\mbox{.}(2024)]%
        {freehand_mechParts_24}
\bibfield{author}{\bibinfo{person}{Zhichao Liao}, \bibinfo{person}{Fengyuan
  Piao}, \bibinfo{person}{Di Huang}, \bibinfo{person}{Xinghui Li},
  \bibinfo{person}{Yue Ma}, \bibinfo{person}{Pingfa Feng},
  \bibinfo{person}{Heming Fang}, {and} \bibinfo{person}{Long Zeng}.}
  \bibinfo{year}{2024}\natexlab{}.
\newblock \showarticletitle{Freehand Sketch Generation from Mechanical
  Components}. In \bibinfo{booktitle}{\emph{Proc. {ACM MM}}}.
  \bibinfo{pages}{6755–6764}.
\newblock


\bibitem[Liu et~al\mbox{.}(2020)]%
        {Liu_2020_CVPR}
\bibfield{author}{\bibinfo{person}{Difan Liu}, \bibinfo{person}{Mohamed
  Nabail}, \bibinfo{person}{Aaron Hertzmann}, {and} \bibinfo{person}{Evangelos
  Kalogerakis}.} \bibinfo{year}{2020}\natexlab{}.
\newblock \showarticletitle{Neural Contours: Learning to Draw Lines from 3D
  Shapes}. In \bibinfo{booktitle}{\emph{IEEE/CVF Conference on Computer Vision
  and Pattern Recognition (CVPR)}}.
\newblock


\bibitem[Liu et~al\mbox{.}(2024)]%
        {LiuSketchDream}
\bibfield{author}{\bibinfo{person}{Feng-Lin Liu}, \bibinfo{person}{Hongbo Fu},
  \bibinfo{person}{Yu-Kun Lai}, {and} \bibinfo{person}{Lin Gao}.}
  \bibinfo{year}{2024}\natexlab{}.
\newblock \showarticletitle{SketchDream: Sketch-based Text-To-3D Generation and
  Editing}.
\newblock \bibinfo{journal}{\emph{ACM Trans. Graph.}} \bibinfo{volume}{43},
  \bibinfo{number}{4}, Article \bibinfo{articleno}{44} (\bibinfo{date}{July}
  \bibinfo{year}{2024}), \bibinfo{numpages}{13}~pages.
\newblock
\showISSN{0730-0301}
\urldef\tempurl%
\url{https://doi.org/10.1145/3658120}
\showDOI{\tempurl}


\bibitem[Liu et~al\mbox{.}(2023)]%
        {liu2023zero1to3}
\bibfield{author}{\bibinfo{person}{Ruoshi Liu}, \bibinfo{person}{Rundi Wu},
  \bibinfo{person}{Basile~Van Hoorick}, \bibinfo{person}{Pavel Tokmakov},
  \bibinfo{person}{Sergey Zakharov}, {and} \bibinfo{person}{Carl Vondrick}.}
  \bibinfo{year}{2023}\natexlab{}.
\newblock \bibinfo{title}{Zero-1-to-3: Zero-shot One Image to 3D Object}.
\newblock
\newblock
\showeprint[arxiv]{2303.11328}~[cs.CV]


\bibitem[Nealen et~al\mbox{.}(2007)]%
        {FiberMesh:2007}
\bibfield{author}{\bibinfo{person}{Andrew Nealen}, \bibinfo{person}{Takeo
  Igarashi}, \bibinfo{person}{Olga Sorkine}, {and} \bibinfo{person}{Marc
  Alexa}.} \bibinfo{year}{2007}\natexlab{}.
\newblock \showarticletitle{{FiberMesh}: Designing Freeform Surfaces with 3{D}
  Curves}.
\newblock \bibinfo{journal}{\emph{ACM Transactions on Graphics (Proceedings of
  ACM SIGGRAPH)}} \bibinfo{volume}{26}, \bibinfo{number}{3}
  (\bibinfo{year}{2007}), \bibinfo{pages}{article no.\ 41}.
\newblock


\bibitem[Nicholls and Kennedy(1995)]%
        {nicholls1995cube}
\bibfield{author}{\bibinfo{person}{Andrea~L Nicholls} {and}
  \bibinfo{person}{John~M Kennedy}.} \bibinfo{year}{1995}\natexlab{}.
\newblock \showarticletitle{Foreshortening in Cube Drawings by Children and
  Adults}.
\newblock \bibinfo{journal}{\emph{Perception}} \bibinfo{volume}{24},
  \bibinfo{number}{12} (\bibinfo{year}{1995}), \bibinfo{pages}{1443--1456}.
\newblock
\urldef\tempurl%
\url{https://doi.org/10.1068/p241443}
\showDOI{\tempurl}
\newblock
\shownote{PMID: 8734543}.


\bibitem[Olsen et~al\mbox{.}(2009)]%
        {OLSEN200985}
\bibfield{author}{\bibinfo{person}{Luke Olsen}, \bibinfo{person}{Faramarz~F.
  Samavati}, \bibinfo{person}{Mario~Costa Sousa}, {and}
  \bibinfo{person}{Joaquim~A. Jorge}.} \bibinfo{year}{2009}\natexlab{}.
\newblock \showarticletitle{Sketch-based modeling: A survey}.
\newblock \bibinfo{journal}{\emph{Computers \& Graphics}} \bibinfo{volume}{33},
  \bibinfo{number}{1} (\bibinfo{year}{2009}), \bibinfo{pages}{85--103}.
\newblock
\showISSN{0097-8493}
\urldef\tempurl%
\url{https://doi.org/10.1016/j.cag.2008.09.013}
\showDOI{\tempurl}


\bibitem[Pepperell and Haertel(2014)]%
        {Pepperell2014}
\bibfield{author}{\bibinfo{person}{Robert Pepperell} {and}
  \bibinfo{person}{Manuela Haertel}.} \bibinfo{year}{2014}\natexlab{}.
\newblock \showarticletitle{Do Artists Use Linear Perspective to Depict Visual
  Space?}
\newblock \bibinfo{journal}{\emph{Perception}} \bibinfo{volume}{43},
  \bibinfo{number}{5} (\bibinfo{year}{2014}), \bibinfo{pages}{395--416}.
\newblock


\bibitem[Reith and Liu(1995)]%
        {reith1995projective}
\bibfield{author}{\bibinfo{person}{Emiel Reith} {and}
  \bibinfo{person}{Chang~Hong Liu}.} \bibinfo{year}{1995}\natexlab{}.
\newblock \showarticletitle{What Hinders Accurate Depiction of Projective
  Shape?}
\newblock \bibinfo{journal}{\emph{Perception}} \bibinfo{volume}{24},
  \bibinfo{number}{9} (\bibinfo{year}{1995}), \bibinfo{pages}{995--1010}.
\newblock
\urldef\tempurl%
\url{https://doi.org/10.1068/p240995}
\showDOI{\tempurl}
\newblock
\shownote{PMID: 8552463}.


\bibitem[Schmidt et~al\mbox{.}(2009a)]%
        {singh_perspective:09}
\bibfield{author}{\bibinfo{person}{Ryan Schmidt}, \bibinfo{person}{Azam Khan},
  \bibinfo{person}{Gord Kurtenbach}, {and} \bibinfo{person}{Karan Singh}.}
  \bibinfo{year}{2009}\natexlab{a}.
\newblock \showarticletitle{On expert performance in 3D curve-drawing tasks}.
  In \bibinfo{booktitle}{\emph{SBIM}}. \bibinfo{address}{New York, NY, USA},
  \bibinfo{pages}{133–140}.
\newblock


\bibitem[Schmidt et~al\mbox{.}(2009b)]%
        {schmidt2009scaffold}
\bibfield{author}{\bibinfo{person}{Ryan Schmidt}, \bibinfo{person}{Azam Khan},
  \bibinfo{person}{Karan Singh}, {and} \bibinfo{person}{Gord Kurtenbach}.}
  \bibinfo{year}{2009}\natexlab{b}.
\newblock \showarticletitle{Analytic drawing of 3D scaffolds}.
\newblock \bibinfo{journal}{\emph{ACM Trans. Graph.}} \bibinfo{volume}{28},
  \bibinfo{number}{5} (\bibinfo{date}{Dec.} \bibinfo{year}{2009}),
  \bibinfo{pages}{1–10}.
\newblock
\showISSN{0730-0301}
\urldef\tempurl%
\url{https://doi.org/10.1145/1618452.1618495}
\showDOI{\tempurl}


\bibitem[Shao et~al\mbox{.}(2012)]%
        {crossShade:12}
\bibfield{author}{\bibinfo{person}{Cloud Shao}, \bibinfo{person}{Adrien
  Bousseau}, \bibinfo{person}{Alla Sheffer}, {and} \bibinfo{person}{Karan
  Singh}.} \bibinfo{year}{2012}\natexlab{}.
\newblock \showarticletitle{CrossShade: shading concept sketches using
  cross-section curves}.
\newblock \bibinfo{journal}{\emph{ACM Trans. Graph.}} \bibinfo{volume}{31},
  \bibinfo{number}{4}, Article \bibinfo{articleno}{45} (\bibinfo{date}{July}
  \bibinfo{year}{2012}), \bibinfo{numpages}{11}~pages.
\newblock


\bibitem[Singh(2002)]%
        {Singh:2002:FP}
\bibfield{author}{\bibinfo{person}{Karan Singh}.}
  \bibinfo{year}{2002}\natexlab{}.
\newblock \showarticletitle{A Fresh Perspective}. In
  \bibinfo{booktitle}{\emph{Proceedings - Graphics Interface}}.
\newblock


\bibitem[Steadman(2002)]%
        {steadman2002vermeer}
\bibfield{author}{\bibinfo{person}{Philip Steadman}.}
  \bibinfo{year}{2002}\natexlab{}.
\newblock \bibinfo{booktitle}{\emph{Vermeer's camera: uncovering the truth
  behind the masterpieces}}.
\newblock \bibinfo{publisher}{Oxford University Press}.
\newblock


\bibitem[Sutherland(1964)]%
        {Sutherland}
\bibfield{author}{\bibinfo{person}{Ivan~E Sutherland}.}
  \bibinfo{year}{1964}\natexlab{}.
\newblock \showarticletitle{Sketch pad a man-machine graphical communication
  system}. In \bibinfo{booktitle}{\emph{Proceedings of the SHARE design
  automation workshop}}. \bibinfo{pages}{6--329}.
\newblock


\bibitem[Tang et~al\mbox{.}(2023)]%
        {tang2023emergent}
\bibfield{author}{\bibinfo{person}{Luming Tang}, \bibinfo{person}{Menglin Jia},
  \bibinfo{person}{Qianqian Wang}, \bibinfo{person}{Cheng~Perng Phoo}, {and}
  \bibinfo{person}{Bharath Hariharan}.} \bibinfo{year}{2023}\natexlab{}.
\newblock \showarticletitle{Emergent Correspondence from Image Diffusion}. In
  \bibinfo{booktitle}{\emph{Thirty-seventh Conference on Neural Information
  Processing Systems}}.
\newblock
\urldef\tempurl%
\url{https://openreview.net/forum?id=ypOiXjdfnU}
\showURL{%
\tempurl}


\bibitem[Taylor and Mitchell(1997)]%
        {taylor1997judgments}
\bibfield{author}{\bibinfo{person}{Laura~M Taylor} {and} \bibinfo{person}{Peter
  Mitchell}.} \bibinfo{year}{1997}\natexlab{}.
\newblock \showarticletitle{Judgments of apparent shape contaminated by
  knowledge of reality: Viewing circles obliquely}.
\newblock \bibinfo{journal}{\emph{British Journal of Psychology}}
  \bibinfo{volume}{88}, \bibinfo{number}{4} (\bibinfo{year}{1997}),
  \bibinfo{pages}{653--670}.
\newblock


\bibitem[Van~Mossel et~al\mbox{.}(2021)]%
        {strokestrip:21}
\bibfield{author}{\bibinfo{person}{Dave~Pagurek Van~Mossel},
  \bibinfo{person}{Chenxi Liu}, \bibinfo{person}{Nicholas Vining},
  \bibinfo{person}{Mikhail Bessmeltsev}, {and} \bibinfo{person}{Alla Sheffer}.}
  \bibinfo{year}{2021}\natexlab{}.
\newblock \showarticletitle{StrokeStrip: joint parameterization and fitting of
  stroke clusters}.
\newblock \bibinfo{journal}{\emph{ACM Trans. Graph.}} \bibinfo{volume}{40},
  \bibinfo{number}{4}, Article \bibinfo{articleno}{50} (\bibinfo{date}{July}
  \bibinfo{year}{2021}), \bibinfo{numpages}{18}~pages.
\newblock


\bibitem[Vinker et~al\mbox{.}(2022)]%
        {vinker2022clipasso}
\bibfield{author}{\bibinfo{person}{Yael Vinker}, \bibinfo{person}{Ehsan
  Pajouheshgar}, \bibinfo{person}{Jessica~Y. Bo},
  \bibinfo{person}{Roman~Christian Bachmann}, \bibinfo{person}{Amit~Haim
  Bermano}, \bibinfo{person}{Daniel Cohen-Or}, \bibinfo{person}{Amir Zamir},
  {and} \bibinfo{person}{Ariel Shamir}.} \bibinfo{year}{2022}\natexlab{}.
\newblock \showarticletitle{CLIPasso: Semantically-Aware Object Sketching}.
\newblock \bibinfo{journal}{\emph{ACM Trans. Graph.}} \bibinfo{volume}{41},
  \bibinfo{number}{4}, Article \bibinfo{articleno}{86} (\bibinfo{date}{jul}
  \bibinfo{year}{2022}), \bibinfo{numpages}{11}~pages.
\newblock


\bibitem[Viterbi(1967)]%
        {viterbi}
\bibfield{author}{\bibinfo{person}{A. Viterbi}.}
  \bibinfo{year}{1967}\natexlab{}.
\newblock \showarticletitle{Error bounds for convolutional codes and an
  asymptotically optimum decoding algorithm}.
\newblock \bibinfo{journal}{\emph{IEEE Transactions on Information Theory}}
  \bibinfo{volume}{13}, \bibinfo{number}{2} (\bibinfo{year}{1967}),
  \bibinfo{pages}{260--269}.
\newblock
\urldef\tempurl%
\url{https://doi.org/10.1109/TIT.1967.1054010}
\showDOI{\tempurl}


\bibitem[Wang et~al\mbox{.}(2024)]%
        {wang2024instantstyle}
\bibfield{author}{\bibinfo{person}{Haofan Wang}, \bibinfo{person}{Peng Xing},
  \bibinfo{person}{Renyuan Huang}, \bibinfo{person}{Hao Ai},
  \bibinfo{person}{Qixun Wang}, {and} \bibinfo{person}{Xu Bai}.}
  \bibinfo{year}{2024}\natexlab{}.
\newblock \showarticletitle{InstantStyle-Plus: Style Transfer with
  Content-Preserving in Text-to-Image Generation}.
\newblock \bibinfo{journal}{\emph{arXiv preprint arXiv:2407.00788}}
  (\bibinfo{year}{2024}).
\newblock


\bibitem[Xiao et~al\mbox{.}(2022)]%
        {xiao2022differsketching}
\bibfield{author}{\bibinfo{person}{Chufeng Xiao}, \bibinfo{person}{Wanchao Su},
  \bibinfo{person}{Jing Liao}, \bibinfo{person}{Zhouhui Lian},
  \bibinfo{person}{Yi-Zhe Song}, {and} \bibinfo{person}{Hongbo Fu}.}
  \bibinfo{year}{2022}\natexlab{}.
\newblock \showarticletitle{DifferSketching: How Differently Do People Sketch
  3D Objects?}
\newblock \bibinfo{journal}{\emph{{ACM SIGGRAPH Asia}}} \bibinfo{volume}{41},
  \bibinfo{number}{4} (\bibinfo{year}{2022}), \bibinfo{pages}{1--16}.
\newblock


\bibitem[Xu et~al\mbox{.}(2014)]%
        {true2Form:2014}
\bibfield{author}{\bibinfo{person}{Baoxuan Xu}, \bibinfo{person}{William
  Chang}, \bibinfo{person}{Alla Sheffer}, \bibinfo{person}{Adrien Bousseau},
  \bibinfo{person}{James McCrae}, {and} \bibinfo{person}{Karan Singh}.}
  \bibinfo{year}{2014}\natexlab{}.
\newblock \showarticletitle{True2Form: 3D curve networks from 2D sketches via
  selective regularization}.
\newblock \bibinfo{journal}{\emph{ACM Trans. Graph.}} \bibinfo{volume}{33},
  \bibinfo{number}{4}, Article \bibinfo{articleno}{131} (\bibinfo{date}{July}
  \bibinfo{year}{2014}), \bibinfo{numpages}{13}~pages.
\newblock


\bibitem[Yoon(2009)]%
        {matchingHMM:09}
\bibfield{author}{\bibinfo{person}{Byung-Jun Yoon}.}
  \bibinfo{year}{2009}\natexlab{}.
\newblock \showarticletitle{Hidden Markov Models and their Applications in
  Biological Sequence Analysis}.
\newblock \bibinfo{journal}{\emph{Current genomics}}  \bibinfo{volume}{10}
  (\bibinfo{date}{09} \bibinfo{year}{2009}), \bibinfo{pages}{402--15}.
\newblock
\urldef\tempurl%
\url{https://doi.org/10.2174/138920209789177575}
\showDOI{\tempurl}


\bibitem[Zhang et~al\mbox{.}(2022)]%
        {zhang2022creatureshop}
\bibfield{author}{\bibinfo{person}{Congyi Zhang}, \bibinfo{person}{Lei Yang},
  \bibinfo{person}{Nenglun Chen}, \bibinfo{person}{Nicholas Vining},
  \bibinfo{person}{Alla Sheffer}, \bibinfo{person}{Francis~C.M. Lau},
  \bibinfo{person}{Guoping Wang}, {and} \bibinfo{person}{Wenping Wang}.}
  \bibinfo{year}{2022}\natexlab{}.
\newblock \showarticletitle{CreatureShop: Interactive 3D Character Modeling and
  Texturing from a Single Color Drawing}.
\newblock \bibinfo{journal}{\emph{IEEE Transactions on Visualization and
  Computer Graphics}} (\bibinfo{year}{2022}), \bibinfo{pages}{1--18}.
\newblock


\end{thebibliography}
